\documentclass[12pt,preprint]{aastex}
\newcommand{\Lya}{\ensuremath{\hbox{Ly}\alpha~}}
\newcommand{\be}{\begin{equation}}
\newcommand{\bea}{\begin{eqnarray}}
\newcommand{\ee}{\end{equation}}
\newcommand{\eea}{\end{eqnarray}}
\newcommand{\A}{\AA~}
\def\HI{\ion{H}{1}}

\def\HI{\ion{H}{1}}
\def\HII{\ion{H}{2}}
\def\hMpc{$h^{-1}$Mpc}

\begin{document}

\title{ Ly$\alpha$ Leaks in the Absorption Spectra of High Redshift QSOs }

\author{Jiren Liu\altaffilmark{1}, Hongguang
Bi\altaffilmark{2,3}, and Li-Zhi Fang\altaffilmark{1}}

\altaffiltext{1}{Department of Physics, University of Arizona,
Tucson, AZ 85721}

\altaffiltext{2}{Purple Mountain Observatory, Nanjing, 210008, P.R.
China.}

\altaffiltext{3}{National Astronomical Observatories, Chinese
Academy of
    Science, Chao-Yang District, Beijing 100012, China}

\begin{abstract}

Spectra of high redshift QSOs show deep Gunn-Peterson absorptions
on the blue sides of the \Lya emissions lines. They can be
decomposed into components called \Lya leaks, defined to be emissive
regions in complementary to otherwise zero-fluxed absorption gaps.
Just like \Lya absorption forests at low redshifts, \Lya leaks are both
easy to find in observations and containing rich sets of statistical
properties that can be used to study the early evolution of the IGM.
Among all properties of a leak profile, we investigate its equivalent
width in this paper, since it is weakly affected by instrumental
resolution and noise. Using 10 Keck QSO spectra at $z\sim6$, we have measured
the number density distribution function $n(W,z)$, defined to be the
number of leaks per equivalent width $W$ and per redshift $z$,
in the redshift range $5.4 - 6.0$. These new observational statistics,
in both the differential and cumulative forms, fit well to hydro numerical
simulations of uniform ionizing background in the $\Lambda$CDM cosmology.
In this model, Ly $\alpha$ leaks are mainly due to low density voids.
It supports the early studies that the IGM at $z\simeq6$ would still be in
a highly ionized state with neutral hydrogen fraction $\simeq 10^{-4}$.
Measurements of $n(W,z)$ at $z>6$ would be effective to probe the 
reionization of the IGM.

\end{abstract}


\section{Introduction}

The absorption spectra of QSOs at low redshift show \Lya forests,
which have played an important role to understand the physical
status of diffused cosmic baryon gas and the ionizing background. At
redshift $z > 5$, however, they are no longer to show forests
features, but consist of complete absorption troughs separated by
spikes of the transmitted flux (e.g. Becker et al. 2001; Fan et al.
2006). That is, although the cosmic hydrogen gas at $z>5$, in
average, is opaque for Ly$\alpha$ photons, there are many tiny
regions, which are Gunn-Peterson transparent and lead to Ly$\alpha$
photon leaking.

The nature of the leaking is crucial to understand the physics of
reionization. According to commonly accepted scenario of
reionization, at early stage, only isolated patches around ionizing
sources are highly ionized. The subsequent growing and overlapping
of the ionizing patches lead to a uniform ionizing background and
the end of reionization (e.g., Ciardi et al. 2003; Sokasian et al.
2003; Gnedin 2004; Mellema et al. 2006) The ionization fraction of
the IGM and the ionizing radiation underwent an evolution from
highly non-uniform patches to a quasi-homogeneous field. Before the
patch-to-uniform transition, only ionized patches would be
transparent to \Lya photons. After the transition, the low density
voids will also be Gunn-Peterson transparent.  Therefore, the origin
of Ly$\alpha$ leaks will constrain the epoch of the patch-to-uniform
transition. In this {\it Letter}, we study the origin of \Lya leaks
in the observed spectra of QSOs at $z\simeq6$.

Several statistics have been introduced to describe the
transmitted flux of Ly$\alpha$ absorption at high redshifts,
including the probability distribution function (PDF) of the flux
(Fan et al. 2002; Becker et al. 2007), the distribution of the size
of dark gaps (Songaila \& Cowie 2002; Fan et al. 2006), and the
largest peak width distribution (Gallerani et al. 2007). We will
focus on the profile of the leaking features in the transmitted
flux. We fit these statistical features with samples of
hydrodynamic simulation with a uniform ionizing background, and
analyze the possibility of explaining the leaks by ionized patches
embedded in neural IGM background.

\section{Samples}

1. {\it  Observational spectra of high redshift QSOs.} The
observational spectra used here are 10 of the 12 Keck spectra of
QSOs at redshift $z>5.8$ compiled in Fan et al. (2006). We excluded
2 BAL QSOs. The data have a uniform resolution of $R\sim4000$ and
are re-binned to a resolution $R=2600$. To avoid the mixing of
Ly$\beta$ absorption and the effect of QSO's \HII~ region, only the
rest frame wavelength between 1050 to 1170 \A are used.
To study the evolution of \Lya leaks, we divide the spectra into
two redshift bins, $5.4<z\leq 5.7$ and $5.7<z\leq 6.0$.
The observed flux, $f_{obs}$, is normalized with a power-law continuum
$f_{con}\propto \nu^{-0.5}$.
The noise level of transmitted flux $F\equiv f_{obs}/f_{con}$ is about
$ 0.018 \pm 0.012$ and $0.014\pm 0.008$ for above two redshift bins.
For more details, we refer to Fan et al. (2006).

2. {\it Simulation samples.} We simulate Ly$\alpha$ absorption
spectra with a hybrid gas/dark matter code based on WENO scheme
(Feng et al. 2004). The simulation is performed in a comoving box of
100\hMpc \ with a $512^3$ mesh. We use the concordance
$\Lambda$CDM cosmology model with parameters to be $\Omega_m$=0.27,
$\Omega_b$=0.044, $\Omega_\Lambda$=0.73, $h$=0.71,
$\sigma_8$=0.84, and spectral index $n=1$.

It has been shown that the observed dramatic decrease and abnormally
large scatter of Gunn-Peterson optical depth at $z\simeq 6$ (Fan et
al. 2006) can be well fitted by models of a uniform ionizing
background (Lidz et al. 2006; Liu et al. 2006). That is, the large
scattering of Gunn-Peterson optical depth may still be mainly due to the
inhomogeneity of the IGM density field. Therefore, we investigate
whether such a uniform ionizing background  can explain the
leaks. In this context, the uniform photoionization rate is
adjusted to yield the same mean optical depth as observational data
at the redshifts considered.  A thermal energy of $T=3\times10^4$K
is added at $z=10$, and only adiabatic cooling and shock heating are
followed. With this method, the photoionization rate
are found to be equal to 0.63 and 0.33 in units of $10^{-12}s^{-1}$
at redshift $z=$5.55 and 5.85, and the corresponding neutral hydrogen
fraction are $3\times10^{-5}$ and $7\times10^{-5}$.
The simulated
spectra were smoothed with a Gaussian window of a FWHM corresponding
to $R\sim4000$, and re-binned to pixels of the size of $R=2600$. We
add Gaussian noises to the re-binned fluxes with variances equal to
the observational noise level.

\section{Statistics of Ly$\alpha$ leaks}

1. {\it Identification of \Lya leaks}.

Ly$\alpha$ leaks are identified as contiguous pixels where the
fluxes have a maximum larger than 2 or 3 $\sigma$ of the local noise
level. The boundaries of a leak are defined as positions where the
fluxes are smaller than a threshold $F_{th}$, or are the minimum
between the neighboring leaks. That is, if there are two local
maximums above 2 or 3 $\sigma$ of the noise level, each one is
identified as a leak. We take the threshold $F_{th}=0.02$ in this
letter. Note the identification of a leak depends mostly on the
condition of the maximum flux (see discussion on Figure 1 below). With
this method we decompose the transmitted fluxes between Gunn-Peterson
troughs into \Lya leaks of different profiles. The \Lya leaks contain
information different from the size of dark gaps and largest peak width,
both of which measure only length scales.

To test the identification condition, we count the number of fake
leaks due to noise in 100 simulation samples. The percentage of fake
leaks are 2.3\%(13\%) and 0.4\%(1.4\%) for the 2 $\sigma$ and
3$\sigma$ identification in redshifts range $5.4-5.7$($5.7-6.0)$.
Similarly, we also count the number of missed leaks due to noise.
The percentage of such missing leaks are 5\%(8\%) and 15\%(22\%) for
the 2(3) $\sigma$ in redshifts range $5.4-5.7$($5.7-6.0)$,
respectively. The fluxes of missing leaks generally are around
$F_{th}$. Therefore, the leak identification with $F_{th}=0.02$ is
statistically reliable. In the 10 Keck spectra, there are totally
173 and 147 leaks in the redshift range $5.4 \leq z \leq 5.7$, and
39 and 32 leaks in $5.7< z \leq 6.0$ for the 2 and 3 $\sigma$
identification, respectively. The fluxes of smallest leaks are a
little higher than $F=0.02$, while big leaks can have $F\simeq 0.3$.

2. {\it Equivalent width functions}.

Similar to emission and absorption lines, we can measure the profile
of \Lya leak with equivalent width, which is defined as the area
under its flux profile, $W=\int Fd\lambda$, where the integral is
over the range between the boundaries. For our observed samples, $W$
lasts in the range from 0.06 \AA \ to about 5 \AA. In general, the
equivalent width $W$ measures the strength of the leaking, or the
Gunn-Peterson optical depth within the leaking regions. The
statistical description we used is the equivalent width function
$n(W,z)$, which is the number of leaks of $W$ at redshift $z$ per unit
$W$ per unit $z$. The equivalent width function reflects the
distribution of the strength of leaking.

We count the observed $W$ into 15 bins with logarithm size $\Delta
\ln W=(1/15)\ln (10/0.01)$. The results are shown in Figure 1, which
is for leaks at redshifts $5.4-5.7$ (top) and $5.7-6.0$ (bottom), and
the 2 $\sigma$ (left) and 3 $\sigma$ (right) identification. The
error bar is of Poisson fluctuation. The functions $n(W,z)$
are weakly dependent on the identification. Although the total
numbers of leaks of 2$\sigma$ and 3$\sigma$ samples are different,
the shape of $n(W,z)$ for both samples are about the same.
As expected, for large leaks of $W>0.5$ \AA, the functions $n(W,z)$
are independent of the 2 or 3$\sigma$ condition, while for
small leaks of $W<0.5$ \AA, the function $n(W,z)$ of 3$\sigma$ is
a little lower than that of 2$\sigma$.

Figure 1 also shows the results given by 100 simulation samples. The
solid curves are the mean of the samples, and the dot lines give the
jackknife error estimator, which is to divide the 100 samples into
10 subsamples and compute the variance over the 10 subsamples. We
see that the distributions of $n(W,z)$ of simulation samples
generally are good to fit the observed samples.
To test the effect of noise, we also calculated the function $n(W,z)$ of
simulation samples without noise addition, and the results are
shown in Figure 1 as the dashed lines. Without noise
addition, the leaks are identified as local maximums above
$F_{th}$=0.02.
 Figure 1 shows that the noise has no effect on big leaks ($W>0.5$ \AA).
 While for small leaks ($W<0.5$ \AA), samples without noise addition
give higher number of leaks than samples with noise addition. This is
because the identification condition of 2 and 3 $\sigma$ is more
rigorous than the condition of  $F_{th}$=0.02. Therefore,  the effect
of noise on $n(W,z)$ does not change the consistence between observed
and modeled $n(W,z)$. The measurement of $W$ is insensitive to
the noise, as it is on the area of profile.

We see from Figure 1 that a few data points at small $W$ show
fluctuation around simulation result. It is probably caused by the
binning. To solve this problem, we calculate the cumulative equivalent
width function defined as $n(>W,z)=\int_{W}^{\infty}n(W,z)dW$,
which is less dependent on the binning.
Since the distributions of leaks of 2 and 3 $\sigma$ identification
are similar, only 2 $\sigma$ identification condition is applied.
The results are presented in Figure 2.
The solid curves are the mean of simulated samples, and the dot lines
give the jackknife error estimator as in Figure 1.
It shows clearly that the cumulative width functions of observed leaks
are smooth, and gives a better fitting with simulation
samples.

Figure 3 presents $n(>W,z)$ vs. $z$ for leaks of $W=0.4, 1$, and
$1.6$\A. The redshift-evolution of leaks with larger $W$ is more
significant than smaller $W$ leaks. This is natural in low density
voids scenario. The larger voids have lower probability, or are the
events given by the tail of the PDF of voids. The PDF tail underwent
a strong evolution at high redshifts. At redshift $z>6$, there are
only very few leaks identified from observation data, and therefore,
we will not extend the analysis to $z>6$.

Since all leaks in simulated spectra are due to low density voids,
the results show that the distribution of observed leaks are
consistent with low density voids assuming the uniform ionizing background.
It is interesting to point out that the tail of the cumulative width
function shown in Figure 2 is close to Gaussian distribution with
respect to logarithm $W$. Therefore, $n(>W,z)$ approximately has a
lognormal tail of $W$.

3. {\it Ionized patches}.

We now estimate the Ly$\alpha$ leaking due to the ionized patches
around ionizing sources. Considering a simple model, ionizing
sources embed in a fully neutral IGM at high redshift. The scale of
ionizing patches can be estimated with radius
$R=R_s[1-\exp(-t/t_{rec})]^{1/3}$, where $R_s$ is the Stromgren sphere
radius, $t_{rec}$ and $t$ are, respectively, the recombination time
and the active age of the ionizing source. It has been shown that,
due to the retardation effect of photon propagation, the scale $R$
actually is an upper limit to the ionized volume (Shapiro et al.
2006; Qiu et al. 2007a). The retardation effect is more apparent for
clustered sources (Qiu et al. 2007b). Moreover, it is also shown that
the fraction of \HI \ within ionized sphere generally is larger
than $10^{-6}$ unless the intensity of sources
$\dot{N}>10^{55}$ sec$^{-1}$ (Qiu et al. 2007a).

It has been shown that the damping wing of the neutral IGM absorption
make ionized patches opaque to \Lya photons if the size is
too small (Miralda-Escude 1998). This effect is more significant if
a small fraction of \HI \ is remained in patches. For instance, an
ionized patch with neutral fraction of $5\times10^{-6}$ around a galaxy
at $z=6$ can yield a flux $F=0.02$ only if the comoving radius $R \geq 3.5$ \hMpc,
or $\dot{N}\geq 9\times10^{53}$ sec$^{-1}$, which requires
a luminosity $L \geq 1.6\times10^{10} L_{\odot}$ if assuming a spectra
of $L_\nu \propto \nu^{-3}$. Here we also assume all the ionizing radiation of
a galaxy is capable to contribute to the ionizing sphere, and the
luminosity $L \geq 1.6\times10^{10} L_{\odot}$ gives a lower limit to
the required luminosity to produce a leak with $F=0.02$.

With these results, one can estimate the number of leaks with $F\geq
0.02$ due to galaxies by using the luminosity function of galaxies
at $z=6$ (Bouwens et al. 2006).  The probability
that a line intercepts patches at an comoving impact parameter
$r=1.5$ \hMpc \ (since the cross radius should be larger than 3.5\hMpc,
we should use a smaller impact radius) for galaxies with luminosity
$>1.6\times10^{10}L_{\odot}$ is (e.g., Peebles 1993)
\bea
\frac{dN}{dz}=
\frac{\pi r^2\phi(>1.6\times10^{10}L_{\odot})c}{H(z)} \sim 7.7,
\eea
here $\phi(>1.6\times10^{10}L_{\odot})$ is the comoving number density of
galaxies with luminosity $>1.6\times10^{10}L_{\odot}$. On the other hand,
Figures 1 and 2 show that the number density of leaks with $F\geq
0.02$ at $5.7<z<6$ is $\simeq 35$. Therefore, if the IGM $z=6$
is mostly neutral, and the only ionized regions are the patches
around galaxies, the leaks of $F\geq 0.02$ given by the ionized
patches of galaxies would be no more than 20\% of the observed
result.

\section{Discussions and conclusions}

The transmitted fluxes between Gunn-Peterson troughs of high
redshift QSO's absorption spectra are not only one peak of the flux,
but contain rich structures, which can be decomposed into \Lya
leaks. The \Lya leaks have profiles similar to emission lines, and
can be measured by equivalent width $W$. The equivalent width
functions, $n(W,z)$ are effective statistical measurement of the process
of reionization. We show that the equivalent width functions
of the observed spectra at redshifts $5.4<z<6.0$ can be well fitted
by hydro simulation of the $\Lambda$CDM cosmology assuming the
ionizing background to be uniform. In this model, all the Ly$\alpha$
leaks are leaking through low density voids.

The mean transmitted flux at $z$ is given by $\bar{F}\propto\int
n(W,z)W dW$. Therefore, by adjusting the photoionization rate to
match the observed Gunn-Peterson optical depth, the mean of $W$ for
simulation samples should be the same with observation. Thus, Figure
1 actually is to show that once we adjusted the mean flux to be the
same as observation, the simulation yields the same distribution of
the observed $W$. In other word, the scattering of $W$ is caused by
the fluctuations of mass density field of HI. Therefore, a small
inhomogeneity of the ionizing background would be allowed. That is,
the distribution of $W$ would still be able to be fitted with a
fluctuated ionizing background if its variance is much less than that
of HI. 

In addition to the distribution of $W$, the evolution of $W$ is                 
also helpful when differentiating models. For example,
in voids scenario, the evolution of $W$ reflects the evolution of low
density voids; while for ionized patches, it reflects the evolution
of the UV luminosity function of ionizing sources.

We show that the ionized patches of galaxies embedded in a fully
neutral IGM at redshift $z\simeq 6$ are not enough to produce the
observed leaks. We also show that leaks can only be produced
by patches around strong ionizing UV photon sources, but not weak sources.
Especially, big leaks ($W>0.5$ \AA, \ or $F>0.1$) have to come from
very strong sources. Therefore, at higher redshift, Ly$\alpha$ leaks
 only probe strong ionizing sources.
Thus, from the existence of many big leaks at $z\leq 6$ we can
conclude that the patch-to-uniform transition of the ionizing
background would occur at $z>6$, and most of the IGM at $z\simeq6$
are still in a highly ionized state of neutral fraction
$f_{HI}\simeq 10^{-4}$. This result is consistent with the analysis
of the transmitted flux PDF (Becker et al. 2007), the QSO proximity
zones (Lidz et al. 2007) and the luminosity function of \Lya
emitting galaxies (Dijkstra et al. 2007).

It should be pointed out that the resolution of the observed data is
low, $\sim3$ \AA, which corresponds to a comoving size $\simeq 0.7$
h$^{-1}$Mpc. In contrast, most of the simulated leaks possess a
intrinsic width $<3$\AA. Thus the low resolution data provide only a
test of smoothed leaking features. One cannot see whether the
smoothed features is due to individual or clustered leaks. Higher
resolution spectra would be able to test not only the width functions,
but also the spatial correlations of the leaks.
They can also provide other measurements of Ly$\alpha$ leaks,
like the FWHM, which would be effective
for a confrontation between real data and models.

The statistics of Ly$\alpha$ leaks at redshifts $\leq 6$ actually
is the statistics of voids formed in the early universe.
The equivalent width functions, $n(W,z)$ of Ly$\alpha$ leaks are similar
to the mass function of galactic clusters. Thus, one can expect
that the width functions of voids are sensitively dependent on
cosmological parameters, and play a similar role
as the mass function of clusters. For instance, the formation of
large voids is found to be sensitively dependent on the mass
parameter $\Omega_m$ (Miranda \& de Araujo 2001). With data of
leaks, we may set constraint on cosmological parameters
at high redshifts. This approach will be reported
separately.

We thank X. Fan for providing the observational spectra and
instructive suggestions. J. L. acknowledges the support of the
International Center for Relativistic Center Network (ICRAnet).
This work is supported in part by the US NSF under the grant
AST-0507340.

\clearpage

\begin{figure}[htb]
\begin{center}
\includegraphics{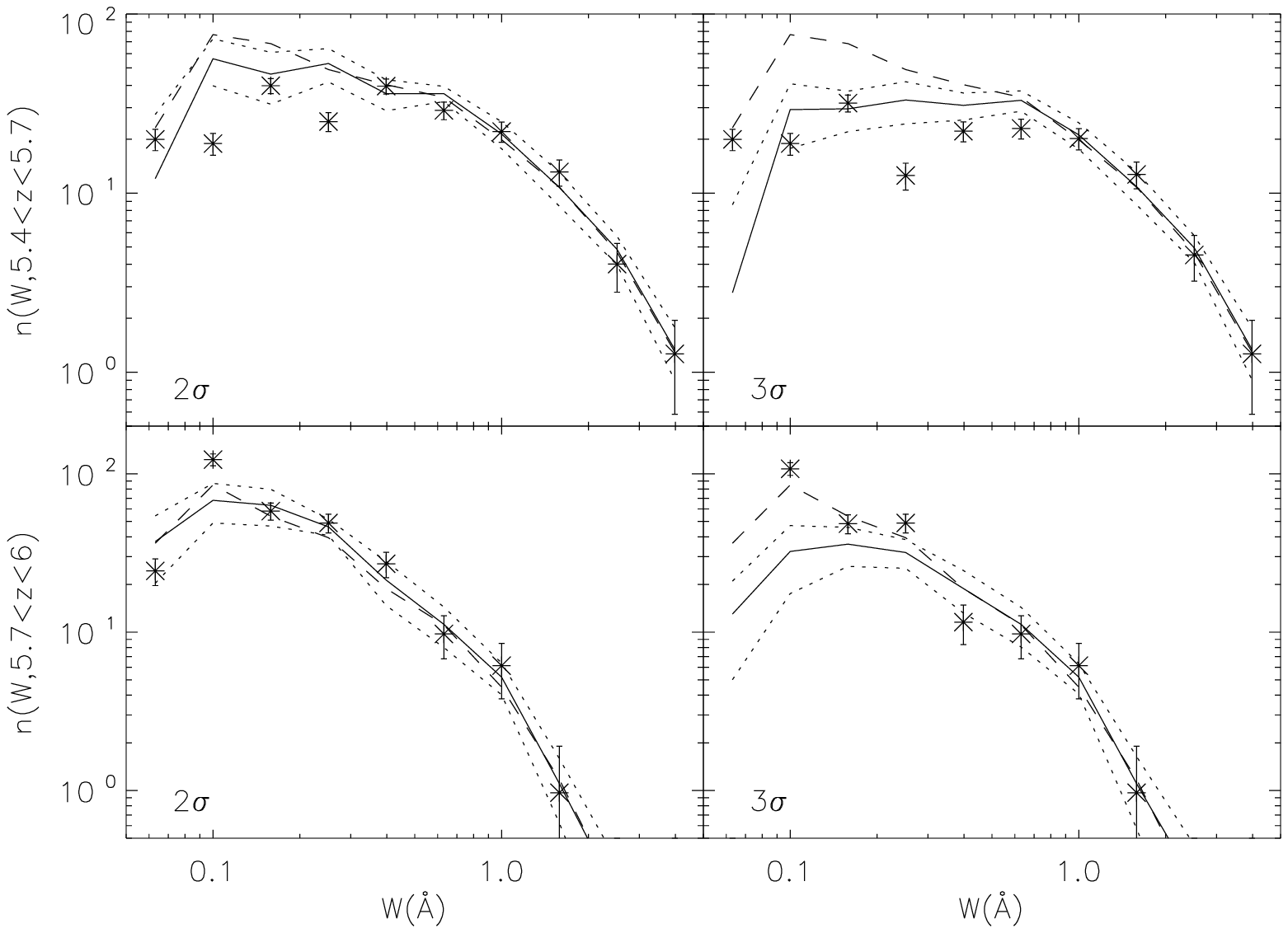}
\caption{Equivalent width function $n(W,z)$ for leaks of 2$\sigma$ (left)
and 3 $\sigma$ (right) identification at redshift ranges $z=5.4-5.7$ (top)
and $5.7-6$ (bottom). The data points are from 10 Keck QSO spectra.
The error bars are from Poisson fluctuation. The solid lines are calculated
with 100
simulated spectra; dot lines are the range of variance over the 10
subsets, each of which contains 10 spectra; dash lines are for samples
without noise. The noise has little effects on equivalent
width $W$.}
\end{center}
\end{figure}

\clearpage

\begin{figure}[htb]
\begin{center}
\includegraphics{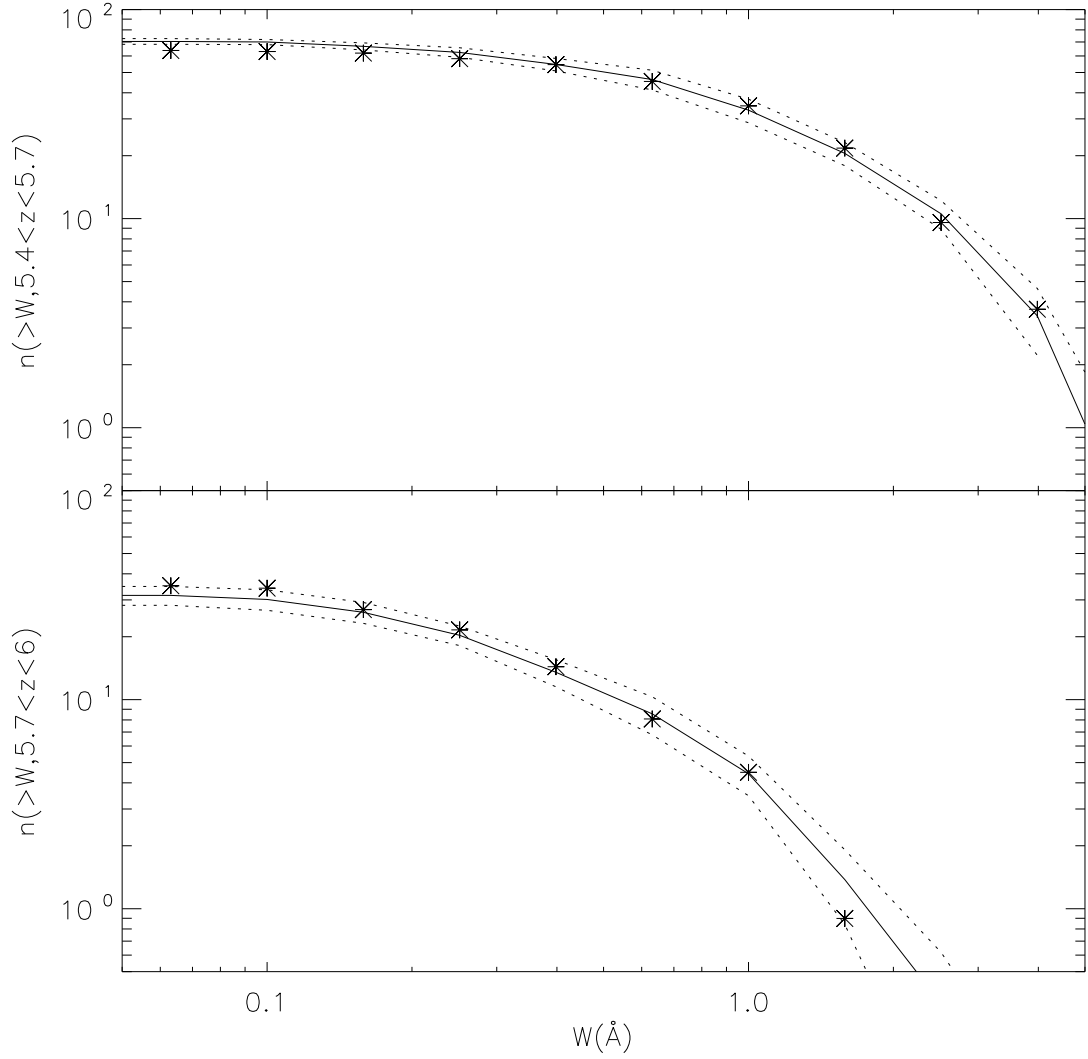}
\caption{Cumulative width function $n(>W,z)$
for leaks at redshift ranges $z=5.4-5.7$ (top) and $5.7-6$ (bottom).
The data points are from 10 Keck QSO spectra.
The solid lines are from 100 simulated
spectra; dot lines are the range of variance over the 10 subsets,
each of which contains 10 spectra. }\label{z6}
\end{center}
\end{figure}

\begin{figure}[htb]
\begin{center}
\includegraphics{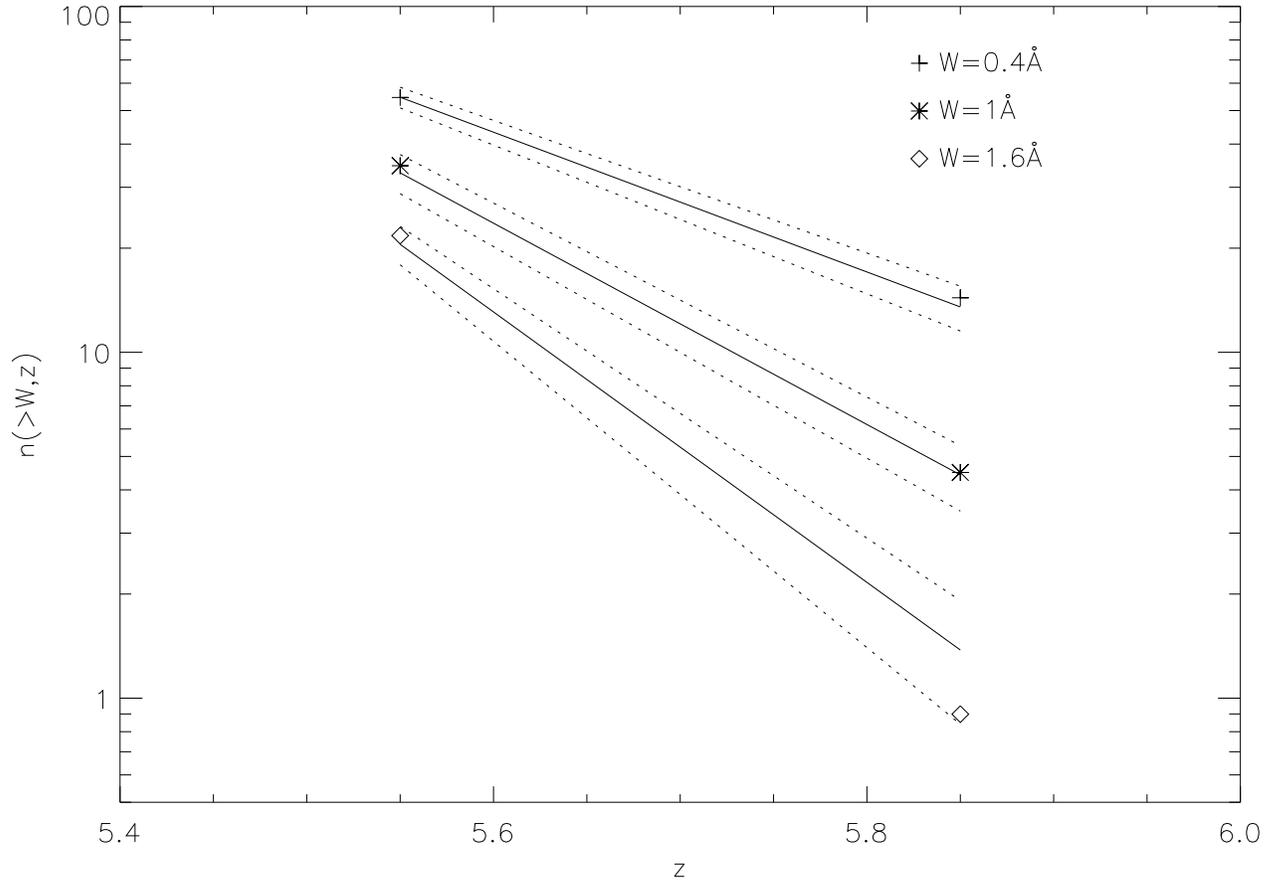}
\caption{The redshift evolution of cumulative width function $n(>W,z)$
of leaks of $W=0.4,1$, and $1.6$\A (top).
The data are taken from Figure 2. The larger leaks evolve faster.
}\label{zz}
\end{center}
\end{figure}

\end{document}